\documentclass[aps,pre,twocolumn,superscriptaddress,groupedaddress]{revtex4-1} 

\usepackage{graphicx}  
\usepackage{dcolumn}   
\usepackage{bm}        
\usepackage{bbold}
\usepackage{appendix}
\usepackage{amssymb, amsmath}   
\usepackage[usenames, dvipsnames]{color}
\usepackage{float}
\usepackage{braket}
\usepackage{import}
\usepackage{pgfplots}

\usepackage{xr-hyper} 
\usepackage{hyperref} 
\hypersetup{
	unicode=true,                                           
	plainpages=false,
	pdffitwindow=true,                                      
	colorlinks=true,                                        
	linkcolor=blue,                                        
	citecolor=blue,                                         
	filecolor=blue,                                        
	urlcolor=blue                                          
}
\pgfplotsset{compat = newest}

\begin{document}
\title{Equivalence of dynamics of disordered quantum ensembles and semi-infinite lattices}
\author{Hallmann Óskar Gestsson}%
\email{hallmann.gestsson.20@ucl.ac.uk}
\author{Charlie Nation}%
\email{c.nation@ucl.ac.uk}
\author{Alexandra Olaya-Castro}%
\email{a.olaya@ucl.ac.uk}
\affiliation{Department of Physics and Astronomy, University College London, London WC1E 6BT, United Kingdom}    

\date{\today}\label{key}

\begin{abstract}

We develop a formalism for mapping the exact dynamics of an ensemble of disordered quantum systems onto the dynamics of a single particle propagating along a semi-infinite lattice, with parameters determined by the probability distribution of disorder realizations of the original heterogeneous quantum ensemble. This mapping provides a geometric interpretation on the loss of coherence when averaging over the ensemble and allows computation of the exact dynamics of the entire disordered ensemble in a single simulation. Alternatively, by exploiting the reverse map, one can obtain lattice dynamics by averaging over realisations of disorder. The potential of this equivalence is showcased with examples of the map in both directions: obtaining dephasing of a qubit via mapping to a lattice model, and solving a simple lattice model via taking an average over realizations of disorder of a unit cell.

\end{abstract}

\maketitle

\textit{Introduction.}―A disordered quantum ensemble is composed of quantum systems whose properties vary randomly from one member of the ensemble to another according to some distribution. Predicting observables of such ensembles then necessitates the introduction of stochastic elements into the Hamiltonian description of a quantum system. 
Examples of disordered ensembles are found across a wide range of fundamental and applied scenarios such as condensed matter physics \cite{Anderson1958Mar, Gornyi2005Nov, Paredes2005, Basko2006May, Bermudez2010, Pal2010Nov, Bardarson2012Jul,Breitweiser2020Jan}, quantum gravity and black hole physics \cite{Maldacena2016Nov, Pollack2020Jul, Peng2021Mar}, quantum optics \cite{Schlawin2012Jul, Gilead2015Sep, Bai2016Mar, Lib2022Sep}, statistical physics \cite{Popescu2006Nov, Cotler2017Nov, Nation2018Oct, Foini2019Apr, Nation2020Oct, Dabelow2020}, quantum chemistry \cite{Frost2010, Gelin2021} and quantum biology \cite{Olaya-Castro2008Aug, Kolli2012Nov, Stross2016Nov, Cohn2022Sep}.

When one considers observable properties of an \textit{ensemble} of disordered quantum systems there will be a noticeable dephasing that is incurred, even if every individual system follows a closed system (unitary) quantum dynamics: information regarding coherences is lost in the averaging that is inextricably carried out when an observable is measured. This is a manifestly classical mechanism that is fundamentally distinct from those inducing incoherent dynamics for a quantum system interacting with an environment. Owing to the broad relevance of such phenomena in physical systems, a central issue of interest thus surrounds the description of the ensemble averaged dynamics of disordered quantum systems, and how this may be related to physically relevant processes.

Recent works have established an effective master equation description of the ensemble averaged state to analyze its dynamical evolution \cite{Gneiting2016Mar, Kropf2016Aug, Chen2018Jan, Gneiting2020Jun}, indicating that disordered quantum systems may be analyzed by applying the frameworks developed within the theory of open quantum systems \cite{Breuer2007}. A unitary description of disordered quantum ensembles has been considered for discrete probability distributions \cite{Paredes2005}. Inspired by the chain-mapping technique \cite{Bulla2005Jan, Bulla2008Apr} that has been applied to open quantum systems \cite{Prior2010Jul, Chin2010Sep}, we consider an alternative and versatile approach where the Hamiltonian of the entire ensemble continuum is transformed using a unitary change of basis that leverages the properties of orthogonal polynomials \cite{Ismail2005Nov}. 

In doing so, we show that the dynamical description of the disordered ensemble is unitarily equivalent to a single semi-infinite lattice \cite{Teschl1999Dec} with couplings that are local for physically relevant models of disorder. Our work then allows gaining new physical insights on the mechanisms on the loss of coherence while at the same time establishing new conceptual links between two apparently dissimilar physical contexts: disordered ensembles and lattice dynamics. 

\begin{figure}
    \includegraphics[width=\linewidth]{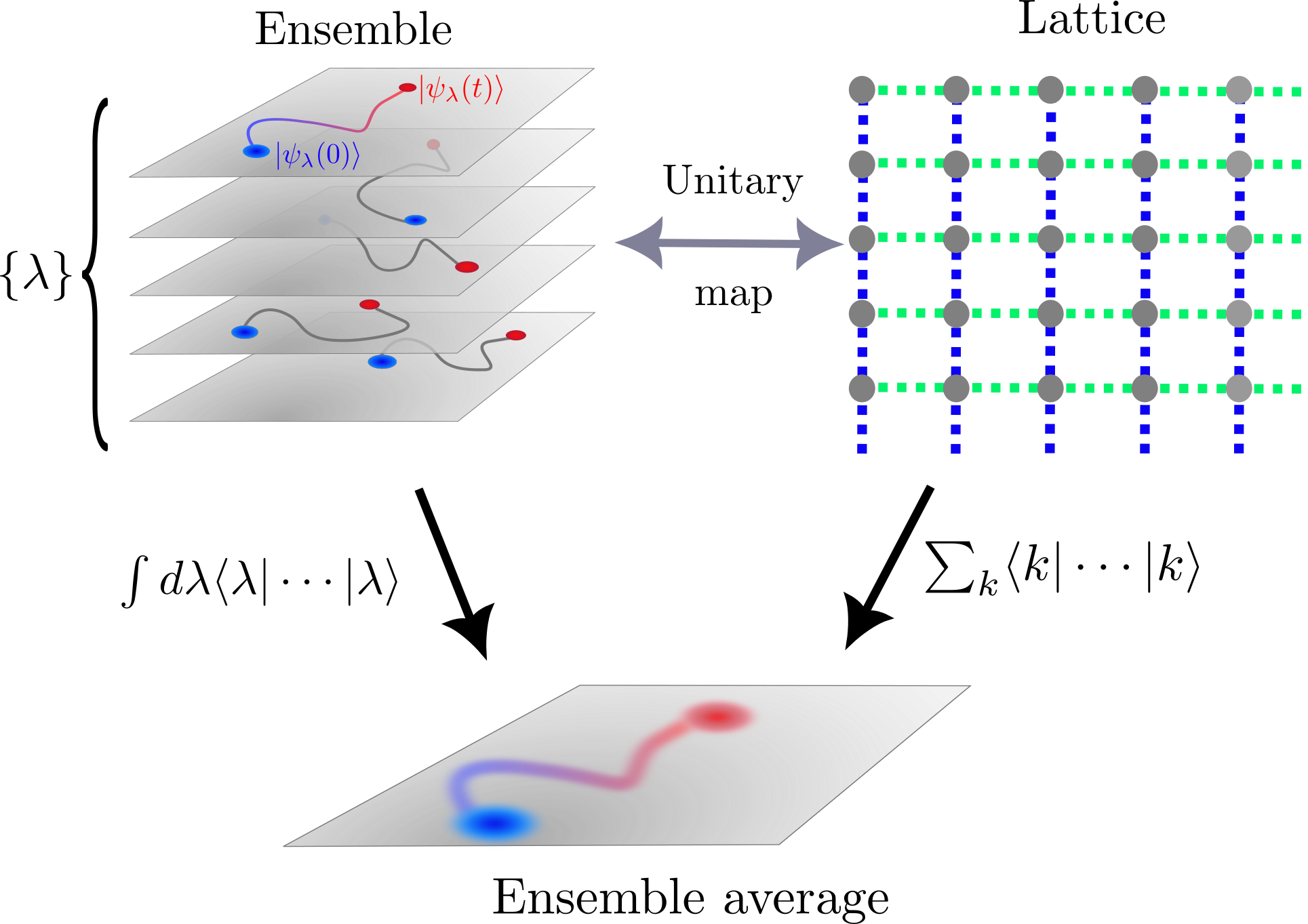}
    \caption{Equivalence of disordered quantum ensembles to lattice dynamics.}
    \label{fig:diagram}
\end{figure}

This lattice representation of the dynamics gives a geometric intuition to the mechanism by which information on coherence is lost when an ensemble average is carried out. Yielding \textit{exact} dynamics of the entire continuum of realizations, which may then be recovered by performing a partial trace over the lattice degree of freedom, bypassing reliance on numerical quadrature rules and sampling from the space of realizations. In practical terms, the advantage of the proposed method stems from the simplicity of the Hamiltonian representation and the accompanying initial wavefunction in the new basis. It is furthermore straightforward to consider initial states of the ensemble that are a function of the disorder, e.g. eigenstates of the system. 

This work is arranged as follows. First, we introduce the pure state formalism for disordered quantum ensembles, and the orthogonal polynomial mapping procedure, showing the general equivalence between the ensemble average dynamics and semi-infinite lattice models. To exemplify our approach, we then discuss the case of linear disorder, and practical aspects of the mapping for numerical application. We then provide examples of the exact equivalence in realistic systems. Additional details are provided in the supplemental material (SM).

\textit{Disordered Quantum Ensembles.}―The ensemble of disordered quantum systems we consider is composed of every possible realization of the disorder, which we assume are not interacting with one another, meaning that each system belonging to the ensemble evolves independent of all other systems in the ensemble. Unitary dynamics of a specific realization of disorder is determined by a Hamiltonian of the form
\begin{equation}\label{Eq:singlerealization}
    \hat{H}_\lambda = \hat{H}_{0} + \hat{V}_\lambda,
\end{equation}
where $\hat{H}_{0}$ and $\hat{V}_\lambda$ are components that are independent and dependent on the disorder, respectively. The multi-index $\lambda$ consists of $l$ unit-less \textit{independent} random variables, such that we have $\lambda = (\lambda_1, \lambda_2,\ldots,\lambda_l)$ and a factorizing joint probability distribution function $p(\lambda) = \prod_{i=1}^l p^{(i)}(\lambda_i)$. 

The state space of each realization is spanned by a basis $\{\ket{n, \lambda}\}_{n=1}^N$ which we use to expand Eq.\ (\ref{Eq:singlerealization}) as
\begin{equation}
    \hat{H}_\lambda = \sum_{n,m=1}^N\left(\braket{n\vert\hat{H}_{0}\vert m} + f_{n,m}(\lambda)\right)\ket{n, \lambda}\bra{m, \lambda},
\end{equation}
where the $\braket{n\vert\hat{H}_{0}\vert m}$ elements are constant for all $\lambda$ so it is dropped as a label, whilst the $f_{n,m}(\lambda) = \braket{n, \lambda\vert\hat{V}_{\lambda}\vert m, \lambda}$ elements vary as a given function of the disorder. We furthermore expand the initial state of each realization with respect to this basis as $\ket{\psi_0, \lambda} = \sum_{n = 1}^N c_n(\lambda)\ket{n, \lambda}$, with $c_n(\lambda) = \braket{n, \lambda\vert\psi_0, \lambda}$. The Hamiltonian of the entire ensemble can be stated in terms of an integral over all possible realizations as
\begin{equation}\label{Eq:IntegralEnsembleHamiltonian}
    \hat{H}_{\text{Ens}} = \int\text{d}\lambda\hat{H}_\lambda,
\end{equation}
with an initial state of the ensemble of the form
\begin{equation}\label{Eq:InitialRealizationBasis}
    \ket{\Psi_0} = \int\text{d}\lambda\sqrt{p(\lambda)}\ket{\psi_0, \lambda}.
\end{equation}
Note that the distribution of disorder informs the initial state of the ensemble (different distributions of disorder result in different initial states) and that $\ket{\Psi_0}$ is normalized as $\braket{\Psi_0\vert\Psi_0} = \int\text{d}\lambda p(\lambda) = 1$. This is a proper physical choice for the initial state of the ensemble as the population density of realizations is given by $p(\lambda)$, such that the probability amplitude may be taken as $\sqrt{p(\lambda)}$. With $\hat{H}_{\text{Ens}}$ and $\ket{\Psi_0}$ at hand we are now in position to determine exact dynamics of the entire ensemble. 

\textit{Recovering ensemble averaged dynamics.}―We now express disorder averaged operators as the partial trace over the disorder label $\lambda$,
\begin{equation}
    \Bar{O}(t) = \int\text{d}\lambda\braket{\lambda\vert O(t)\vert\lambda},
\end{equation}
where $O(t)$ is a generic operator onto the Hilbert space of the entire ensemble, and $\Bar{O}(t)$ is its corresponding disorder average. In particular we have that for average density matrix dynamics with its initial state to be of the form given in Eq.\ (\ref{Eq:InitialRealizationBasis}) we have
\begin{align}
    \Bar{\rho}(t) &= \int\text{d}\lambda\braket{\lambda\vert e^{-i\hat{H}_{\text{Ens}}}\vert\Psi_0}\braket{\Psi_0\vert e^{i\hat{H}_{\text{Ens}}}\vert\lambda} \\
    &= \int\text{d}\lambda p(\lambda) \braket{\lambda\vert e^{-i\hat{H}_\lambda t}\vert\psi_0, \lambda}\braket{\psi_0, \lambda\vert e^{i\hat{H}_\lambda t}\vert\lambda},
\end{align}
where we have made use of the fact that realizations are mutually independent such that $\braket{\lambda\vert e^{-i\hat{H}_{\text{Ens}}}\vert \psi_0,\lambda^\prime} = \braket{\lambda\vert e^{-i\hat{H}_\lambda t}\vert\psi_0, \lambda}\delta(\lambda - \lambda^\prime)$. In the special case where each initial state is independent of $\lambda$ we can further simplify the expression for $\Bar{\rho}(t)$ by making use of $\braket{\lambda\vert e^{-i\hat{H}_\lambda t}\vert\psi_0, \lambda} = e^{-i\hat{H}_\lambda t}\ket{\psi_0}$, yielding
\begin{equation}
    \Bar{\rho}(t) = \int\text{d}\lambda p(\lambda) e^{-i\hat{H}_\lambda t}\ket{\psi_0}\bra{\psi_0}e^{i\hat{H}_\lambda t},
\end{equation}
such that we are able to recover from our pure state formalism the average density matrix dynamics considered in Refs.\ \cite{Gneiting2016Mar, Kropf2016Aug, Chen2018Jan, Gneiting2020Jun}. 

\textit{Mapping ensemble dynamics to a semi-infinite lattice.}―The typical strategy for calculating average dynamics of a disordered quantum ensemble is to sample the space of realizations, either randomly or according to numerical quadratures, which can be parallelized in a straightforward manner as $\hat{H}_{\text{Ens}}$ is block diagonal in $\lambda$. We consider a different approach and perform a change of basis using a family of orthonormal polynomials that transforms the integral representation of Eq.\ (\ref{Eq:IntegralEnsembleHamiltonian}) into a discrete semi-infinite lattice with nearest-neighbour hopping terms. This procedure is equivalent to a Lanczos tridiagonalization of the ensemble continuum Hamiltonian when the disorder parameter enters as a linear term into the Hamiltonian \cite{deVega2015Oct}. Note this approach is also used in the study of operator complexity growth \cite{Parker2019Oct, Muck2022Nov}.

We now introduce the set of polynomials, $\{\phi_k(\lambda)\}_{k=0}^\infty$, that are mutually orthonormal with respect to the measure $p(\lambda)\text{d}\lambda$ (see 
 SM for a brief introduction to orthogonal polynomials, or e.g. Ref.\ \cite{Ismail2005Nov}) and define a new discrete basis,
\begin{equation}\label{Eq:ChangeOfBasis}
    \ket{n, \lambda} = \sum_{k=0}^\infty \sqrt{p(\lambda)}\phi_k(\lambda)\ket{n, k}.
\end{equation}
The disorder has been assumed to consist of $l$ independent random variables such that we have the factorization $p(\lambda) = \prod_{i=1}^lp^{(i)}(\lambda_i)$, which results in a factorization of the polynomials as $\phi_K(\lambda) = \prod_{i=1}^l\phi_{k_i}(\lambda_i)$, where we have introduced a discrete multi-index $K = (k_1, k_2, \ldots, k_l)$. The disorder-independent component of the ensemble Hamiltonian in the discrete basis will reduce as
\begin{equation}\label{Eq:TrivialChangeOfBasis}
    \int\text{d}\lambda\ket{n, \lambda}\bra{m, \lambda} = \sum_K\ket{n, K}\bra{m, K}.
\end{equation}
As for the disorder-dependent terms we have
\begin{equation}
    \int\text{d}\lambda f_{n,m}(\lambda)\ket{n, \lambda}\bra{m, \lambda} = \sum_{K,K^\prime} f_{n,m}^{(K,K^\prime)}\ket{n, K}\bra{m, K^\prime},
\end{equation}
where $f_{n,m}^{(K,K^\prime)} = \int\text{d}\lambda p(\lambda)f_{n,m}(\lambda)\phi_K(\lambda)\phi_{K^\prime}(\lambda)$, such that the Hamiltonian of the ensemble in the lattice basis is written as
\begin{equation}\label{Eq:SemiLatticeHamiltonian}
    \hat{H}_{\text{Ens}} = \sum_{K,K^\prime}\left(\hat{H}_{0,K}\delta_{K,K^\prime} + \sum_{n,m=1}^Nf_{n,m}^{(K,K^\prime)}\ket{n, K}\bra{m, K^\prime}\right),
\end{equation}
with $\hat{H}_{0,K} = \sum_{n,m=1}^N\braket{n\vert\hat{H}_{0}\vert m}\ket{n, K}\bra{m, K}$ and $\delta_{K,K^\prime} = \prod_{i = 1}^l\delta_{k_i,k^\prime_i}$. The SM provides additional details on how one may arrive at Eq.\ (\ref{Eq:SemiLatticeHamiltonian}). The disorder independent components have been projected onto each node of the lattice, whilst disorder dependent terms result in inter- and intra-node coupling terms. 

Therefore the dynamics of the ensemble of Hamiltonians is unitarily equivalent to a single semi-infinite lattice whose dimension is equal to the number of disorder parameters $l$.  We may therefore bypass numerical quadrature techniques that yield approximate solutions to the ensemble dynamics of Eq.\ (\ref{Eq:IntegralEnsembleHamiltonian}) and whose accuracy generally relies on the number of points sampled on the integral interval, and directly compute the exact dynamics of the ensemble via straightforward simulation of the unitary dynamics corresponding to Eq.\ (\ref{Eq:SemiLatticeHamiltonian}). Utility of this change of basis will however rely on both the initial state of the ensemble and the form of the $f_{n,m}(\lambda)$ functions. Note that a disordered Hamiltonian with many random variables will generally result in a high dimensional lattice which can become numerically intractable, such that randomly sampling the space of realizations would be the practical numerical approach for computing average dynamics in such cases. We are able to effectively simulate the exact ensemble dynamics corresponding to low dimensional lattices by utilizing the sparsity of the Hamiltonian matrix representation. 

The initial state of the ensemble can be expanded in terms of the lattice basis states as
\begin{equation}\label{Eq:InitialStateEnsemble_discrete}
    \ket{\Psi_0} = \sum_{n = 1}^N \sum_K d_{n,K} \ket{n, K},
\end{equation}
with $d_{n,K} = \int\text{d}\lambda p(\lambda)\phi_K(\lambda)c_n(\lambda)$ that also serve as expansion coefficients for $c_n(\lambda) = \sum_K d_{n,K} \phi_K(\lambda)$. A special case of the initial state of the ensemble is when the wavefunction of each realization of disorder does \textit{not} change as a function of the disorder, i.e. $c_n(\lambda) = c_n$ for all $\lambda$. We then simply have $c_n(\lambda) = c_n \phi_{\underline{0}}(\lambda)$ which results in Eq.\ (\ref{Eq:InitialStateEnsemble_discrete}) reducing to a single term,
\begin{equation}\label{Eq:InitialStateEnsemble_special}
    \ket{\Psi_0} = \sum_{n = 1}^N c_n \ket{n, \underline{0}}.
\end{equation}
The initial state of the ensemble is found to be entirely localized to the single node at the origin of the lattice due to the distributions of the disorder and initial states being identical in this special case, which yields a simple representation of the initial ensemble wavefunction in the lattice basis. This fact motivates a straightforward truncation scheme to the numerical simulation of the exact ensemble dynamics by introducing a cutoff to the lattice at a distance $D$ from the origin such that the ensemble dynamics up to a time $t < MD$ is accurately simulated, where $M$ is some positive constant determined by the specifics of the disorder model. 

Another form of the initial state of general interest is to consider an ensemble of eigenstates of $\hat{H}_\lambda$, e.g. an ensemble of energy ground states. As each realization is found in a stationary state this case reduces to a case of spectral disorder where the ensemble distribution is determined by the ground state energy distribution $p(E_G(\lambda))$, rather than $p(\lambda)$. The dephasing due to the ensemble average can then be exactly simulated by performing the change of basis, Eq.\ (\ref{Eq:ChangeOfBasis}), with respect to $p(E_G(\lambda))$.

\textit{Linear disorder.}―A ubiquitous functional form of the disorder component is a static linear function $f_{n,m}(\lambda) = \sum_{i=1}^l c_{n,m}^{(i)}\lambda_i$, where $c_{n,m}^{(i)}$ are arbitrary coefficients that depend on the specifics of the model for disorder \cite{Anderson1958Mar, Pal2010Nov, Bardarson2012Jul, Olaya-Castro2008Aug, Kolli2012Nov, Gneiting2016Mar, Kropf2016Aug, Chen2018Jan, Gneiting2020Jun, Kropf2020Aug}. For this case we find the ensemble dynamics to map onto a lattice with nearest-neighbour interaction terms only. The expression for the expansion coefficients $f_{n,m}^{(K,K^\prime)}$ will simplify considerably and be written in terms of polynomial recurrence coefficients as
\begin{equation}\label{Eq:fCoeffs}
    f_{n,m}^{(K,K^\prime)} = \sum_{i=1}^l c_{n,m}^{(i)}(\alpha_{k_i}\delta_{K,K^\prime} + \sqrt{\beta_{k_i+1}}\delta_{K,K^\prime\pm1_i}),
\end{equation}
where $1_i$ denotes a multi-index which has a single non-zero entry of unity as its $i$-th component (see SM for derivation of Eq.\ (\ref{Eq:fCoeffs})). The interaction of the $l$-lattice Hamiltonian is then reduced to nearest-neighbours and we obtain
\begin{equation}
    \begin{aligned}\label{Eq:linear_disorder_Hamiltonian}
        \hat{H}_{\text{Ens}} & = \sum_K\left(\hat{H}_{0,K}  + \sum_{n,m=1}^N\sum_{i=1}^l c_{n,m}^{(i)}\alpha_{k_i}\ket{n, K}\bra{m, K}\right) \\
        & + \sum_{n,m=1}^N\sum_{i=1}^l c_{n,m}^{(i)}\sqrt{\beta_{k_i+1}}\ket{n, K}\bra{m, K+1_i} + \text{h.c.}
    \end{aligned}
\end{equation}
It is by virtue of the three-term recurrence relation that every set of mutually orthogonal polynomials satisfy that we may write $f_{n,m}^{(K,K^\prime)}$ in terms of their recurrence coefficients (see SM for more details regarding the recurrence relation).

\begin{figure}
    \centering
    \includegraphics[width=\linewidth]{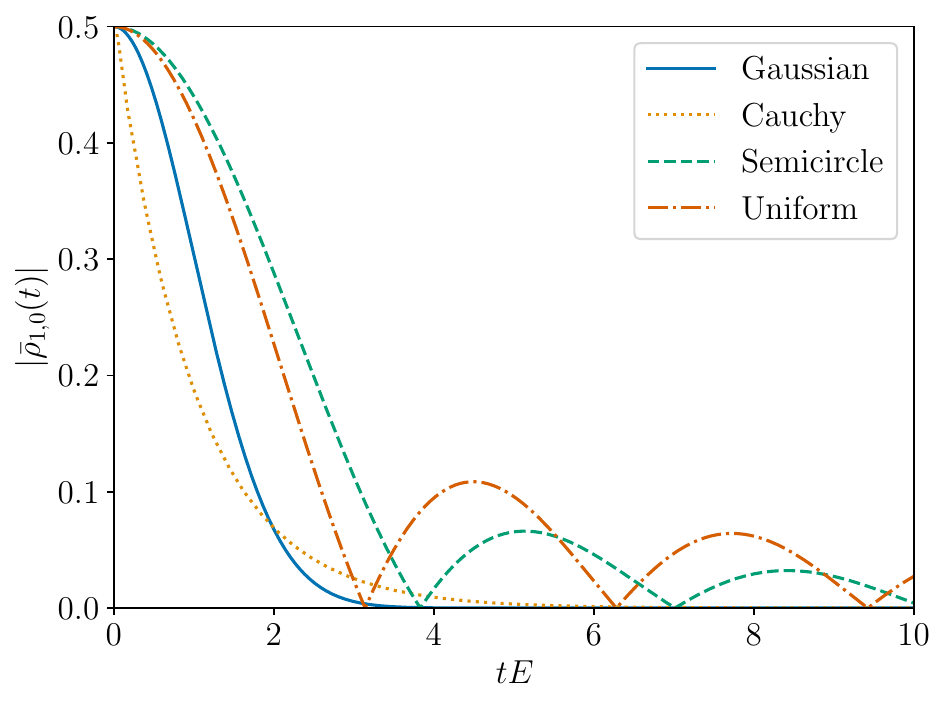}
    \caption{Dephasing dynamics for an ensemble of disordered qubits for different distributions of the energy calculated via simulation of the equivalent lattice model. The form of the distribution has strong influence on how the ensemble will propagate in time. Here we have set $E = E_1 - E_0$. }
    \label{Fig:QubitDephasing}
\end{figure}

\textit{Disordered qubit ensemble dephasing.}―Let us consider an ensemble of non-interacting qubits with randomly sampled excited state energies over a probability distribution $p(\lambda)$, such that each realization is described by a Hamiltonian of the form
\begin{equation}
    \hat{H}_\lambda = E_0\ket{0, \lambda}\bra{0, \lambda} + (E_1 + \lambda)\ket{1, \lambda}\bra{1, \lambda},
\end{equation}
where $E_0$ and $E_1$ are the central energies of the ground and excited state, respectively, and $\lambda$ is the static disorder. The Hamiltonian of the ensemble is then mapped onto a nearest-neighbour coupled chain form via Eq.\ (\ref{Eq:linear_disorder_Hamiltonian}),
\begin{align}
    \hat{H}_{\text{Ens}} &= \sum_{k=0}^\infty\left(E_0\ket{0, k}\bra{0, k} + (E_1 + \alpha_k)\ket{1, k}\bra{1, k}\right)\nonumber \\
    &+ \sum_{k=0}^\infty\sqrt{\beta_{k+1}}\ket{1, k}\bra{1, k+1} + \text{h.c.}
\end{align}
Figure \ref{Fig:QubitDephasing} demonstrates the dephasing dynamics for an initial state where each realization is in the superposition $\ket{\psi_0, \lambda} = (\ket{0, \lambda} + \ket{1, \lambda})/\sqrt{2}$ such that the initial wavefunction of the ensemble may be represented in the lattice basis as $\ket{\Psi_0} = (\ket{0, 0} + \ket{1, 0})/\sqrt{2}$. Dephasing is shown for four different probability distributions: Gaussian, Cauchy, Semicircle, and Uniform. Notice that the semicircle and uniform distributions cases exhibit coherence revivals, which is a consequence of the fact that they have finite support \cite{Kropf2016Aug}. For this particular case it has been found that the dephasing dynamics is characterized completely by the characteristic function of the random variable \cite{Kropf2016Aug} and this can be seen by noting that each realization will have its coherence to be proportional to $e^{i\lambda t}$, and that the expected value of $e^{i\lambda t}$ is the definition of the characteristic function \cite{Lukacs1972Apr}.

We can determine the exact error of our numerical approach for this example as analytical solutions are known. We find that the error is on the order of machine precision where we have considered the Gaussian, semicircle, and uniform distributions, confirming that we are recovering exact dynamics. There is however a finite error for the Cauchy distribution stemming from the energy cutoff we introduced in order to obtain well defined recurrence coefficients. See the SM for details on the parameters used and the closed form solutions we compare to. We also demonstrate in the SM how one may recover the closed form solution of the ensemble averaged dynamics by tracing over the lattice degree of freedom. 

\begin{figure}
    \centering
    \includegraphics[width=\linewidth]{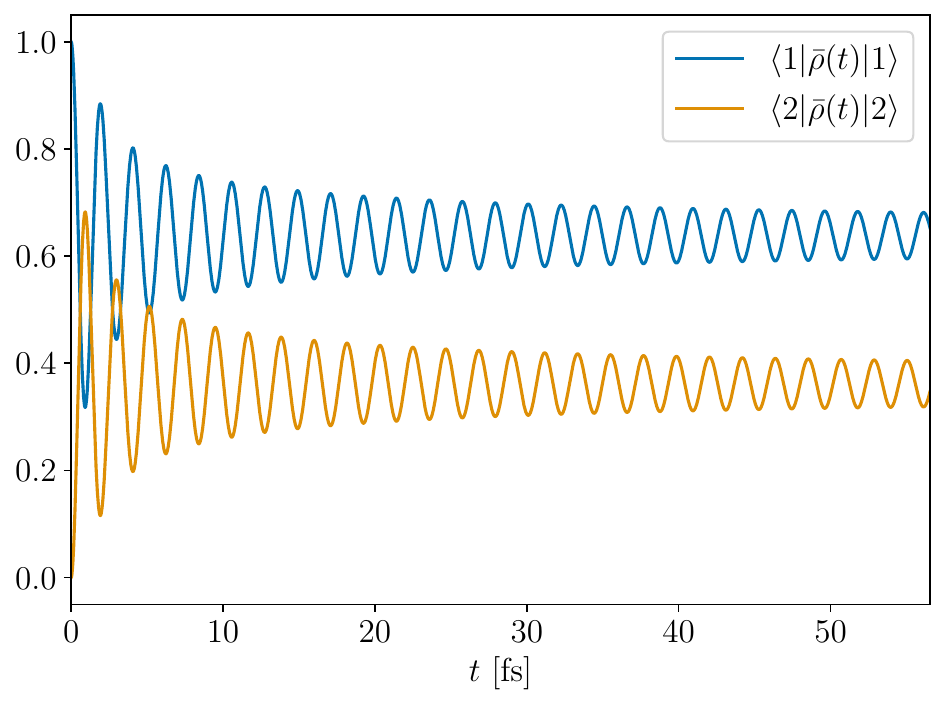}
    \caption{Disorder averaged dynamics for site populations of a dimer sub-unit of the light-harvesting complex LH2 \cite{Cogdell2006Aug}. Novel non-unitary dynamics are exhibited in this example whereby relaxation is incomplete, meaning that the populations reach a new oscillatory state after an initial relaxation. Parameters used are taken from \cite{Novoderezhkin2003Oct}.}
    \label{Fig:DimerPopulations}
\end{figure}

\textit{Disorder averaged dimer dynamics.}―Let us consider an ensemble of non-interacting dimers whose site energies are sampled from independent identical Gaussian distributions, such that each realization is described by a Hamiltonian of the form
\begin{equation}
    \begin{aligned}
        \hat{H}_{\lambda_1,\lambda_2} &= (E_1 + \lambda_1)\ket{1, \lambda_1,\lambda_2}\bra{1, \lambda_1,\lambda_2} \\
        &+\hspace{0.04cm} (E_2 + \lambda_2)\ket{2, \lambda_1,\lambda_2}\bra{2, \lambda_1,\lambda_2} \\
        &+ V(\ket{1, \lambda_1,\lambda_2}\bra{2, \lambda_1,\lambda_2} + \text{h.c.}),
    \end{aligned}
\end{equation}
where $E_1$ and $E_2$ are the two central energies of the site states, $V$ is the coupling, and $\lambda_1$, $\lambda_2$ are the static disorder. The parameters we assume for this particular dimer model are $E_1 = 12325\,\text{cm}^{-1}$, $E_2 = 12025\,\text{cm}^{-1}$, $V = 273\,\text{cm}^{-1}$, so as to simulate a dimerized pair of chromophores \cite{Novoderezhkin2003Oct, Cogdell2006Aug}. The standard deviation we have chosen is $\sigma = 200\,\text{cm}^{-1}$. The change of basis we propose will now expand the ensemble into a two dimensional semi-infinite lattice. Performing the trace over the lattice degrees of freedom will then yield the disorder averaged density matrix in terms of the sites states, whose populations we have plotted in Fig.\ \ref{Fig:DimerPopulations}. We assume that the initial state of the ensemble is such that every instance is localized to site 1. Every instance of the system evolves in time according to the Schrödinger equation, with the population oscillating between sites in a unitary manner. This is not the case for the disordered ensemble average of the dimer as Fig.\ \ref{Fig:DimerPopulations} demonstrates. Here we observe an initial relaxation of the average population dynamics, which is however not a complete relaxation as the average dynamics eventually reach a new oscillating stationary state. 

\textit{Wigner semicircle distribution and constant couplings.}―In the above text we have established an equivalency between disordered quantum ensembles and semi-infinite lattices. When we transform from the $\lambda$ basis and into the $k$ basis, we are effectively leveraging the spectral information we have at hand ($p(\lambda)$) in order to compute the ensemble averaged dynamics. Fully utilizing the duality that we have established, one can go from a Hamiltonian of a semi-infinite lattice with couplings that can be interpreted as recurrence coefficients and (block-) diagonalize the Hamiltonian.

For example, the Wigner semicircle distribution gives rise to recurrence coefficients that are constant and can therefore be used to generate semi-infinite lattices that are translationally invariant for a given unit cell. The corresponding set of orthogonal polynomials will be the Chebyshev polynomials of the second kind, $U_k(\lambda)$, which can be made equivalent to a sine wave basis expansion with a change of variables \cite{Boyd}. By setting $\lambda = \cos\theta$ we have $p(\theta) = \frac{2}{\pi}\sin^2\theta$ and $U_k(\cos\theta) = \frac{\sin((k+1)\theta)}{\sin\theta}$ such that Eq.\ (\ref{Eq:ChangeOfBasis}) will become $\ket{n,\theta} = \sqrt{\frac{2}{\pi}}\sum_{k=0}^\infty \sin((k+1)\theta)\ket{n,k}$.

In other words, the dynamics of a single particle propagating across a semi-infinite constant coupling lattice is equivalent to dynamics of an ensemble whose disorder follows a Wigner semicircle distribution. This stems from the fact that the eigenstates of the lattice form a continuum that is indexed by a quantum number which in turn plays the role of disorder in our ensemble picture. To clarify, what we are highlighting here is not an analytical technique to investigate lattice dynamics, but rather a new interpretation of those dynamics. 

\textit{Conclusions.}―We have shown an equivalency between the dynamics of an ensemble of disordered quantum systems and of a single semi-infinite lattice. This has been achieved by identifying the pure state description of ensembles which may then be unitarily transformed to a lattice description using a basis of orthogonal polynomials. Terms representing disorder in the ensemble picture will correspond to exchange (hopping) terms in the lattice picture. We furthermore show that exact ensemble average dynamics may be recovered by performing a partial trace over the lattice degrees of freedom, yielding non-unitary evolution that is fundamentally distinct from open quantum systems. 

We have demonstrated that the transformation to the lattice picture is exact by simulating disorder-induced dephasing of a qubit and comparing to known analytical solutions. We have also studied the average population dynamics for a disordered dimer which exhibits novel non-unitary dynamics, demonstrating the generality of our approach. Finally, we have discussed the case of semi-infinite lattices with constant nearest-neighbour coupling and highlighted their equivalency to ensembles of quantum systems with disorder that follows a Wigner semicircle distribution. 

\textit{Acknowledgements.}―
We thank the and the Gordon and Betty Moore Foundation (Grant GBMF8820) for financial support and the Engineering and Physical Sciences Research Council (EPSRC UK Grant No. EP/V049011/1). We would like to thank D. Porras for insightful discussions.
\vfill


\pagebreak
\onecolumngrid
\begin{center}
\textbf{\large Supplemental Material: \\ Equivalence of dynamics of disordered quantum ensembles and semi-infinite lattices}
  \vskip 1.0em
  Hallmann Óskar Gestsson, Charlie Nation, and Alexandra Olaya-Castro \\
\small \textit{Department of Physics and Astronomy, University College London, London WC1E 6BT, United Kingdom} \\
  (Dated: \today)
\end{center}
\twocolumngrid

\setcounter{equation}{0}
\setcounter{figure}{0}
\setcounter{table}{0}
\setcounter{page}{1}
\makeatletter

\renewcommand{\theequation}{S\arabic{equation}}
\renewcommand{\thefigure}{S\arabic{figure}}

\section{Primer on orthogonal polynomials}
For a given probability distribution $p$ with finite moments that is defined for the real domain $\mathcal{D} \subseteq \mathbb{R}$ we may define an inner product
\begin{equation}
    \braket{f, g} = \int_\mathcal{D}f(x)g(x)p(x)\text{d}x
\end{equation}
for arbitrary real-valued functions $f$, $g$ defined on $\mathcal{D}$. There then exists a sequence of \textit{monic} orthogonal polynomials $\{P_n(x)\}_{n=0}^\infty$ that satisfies the following condition,
\begin{equation}\label{Eq:MonicCondition}
    \braket{P_n, P_m} = \int_\mathcal{D}P_n(x)P_m(x)p(x)\text{d}x = \zeta_n\delta_{n,m},
\end{equation}
with $\delta_{n,m}$ being the Kronecker delta, and positive constants $\zeta_n$ where $\zeta_0 = 1$ \cite{Ismail2005Nov}. The subscript of the $P_n$ elements is taken to correspond to the degree of the polynomial, e.g. $P_0(x) = 1$. The sequence forms a basis for real-valued functions on $\mathcal{D}$ such that for any $f : \mathcal{D} \rightarrow \mathbb{R}$ there exists a non-zero sequence of real-valued elements $\{c_n\}_{n=0}^\infty$ such that
\begin{equation}\label{Eq:generic_expansion}
    f(x) = \sum_{n=0}^\infty c_nP_n(x),
\end{equation}
where $c_n = \braket{f,P_n} / \zeta_n$. 

We can analogously introduce a corresponding sequence of \textit{orthonormal} polynomials $\{p_n(x)\}_{n=0}^\infty$ whose elements are defined as $p_n(x) = P_n(x) / \sqrt{\zeta_n}$ and will satisfy
\begin{equation}\label{Eq:OrthonormalityCondition}
    \braket{p_n, p_m} = \delta_{n,m}.
\end{equation}

\subsection{Recurrence Relation}
A powerful property of any sequence of orthogonal polynomials, that is at the core of the mapping in the main text, is the three-term recurrence relation they satisfy. For monic polynomials we have
\begin{equation}\label{Eq:ThreeTermRecurrence}
    P_{n+1}(x) = (x - \alpha_n)P_n(x) - \beta_n P_{n-1}(x),
\end{equation}
with initial conditions: $P_0(x) = 1$, $P_1(x) = x - \alpha_0$, and recurrence coefficients $\alpha_n$, $\beta_n$, that are determined as
\begin{equation}\label{Eq:alpha_n_def}
    \alpha_n = \braket{xP_n,P_n} / \braket{P_n,P_n},
\end{equation}
\begin{equation}
    \beta_n = \braket{P_n,P_n} / \braket{P_{n-1},P_{n-1}}.
\end{equation}

We can show that Eq. (\ref{Eq:ThreeTermRecurrence}) holds by expanding $xP_n(x)$ in terms of monic polynomials using Eq.\,\eqref{Eq:generic_expansion},
\begin{equation}\label{Eq:ProofMonicExpansion}
    xP_n(x) = \sum_{k = 0}^{n+1} \frac{\braket{xP_n,P_k}}{\zeta_k}P_k(x),
\end{equation}
where we have noted that the expansion involves only polynomials of degree $n+1$ or lower, as $xP_n(x)$ will be a polynomial of degree $n+1$. Now we consider the coefficients appearing in the expansion and write
\begin{align}
    \braket{xP_n,P_k} & = \braket{P_n,xP_k} \\
    & = \sum_{m = 0}^{k+1}\frac{\braket{P_n,xP_m}}{\zeta_m}\braket{P_n,P_m} \\
    & = \sum_{m = 0}^{k+1}\braket{P_n,xP_m}\delta_{n,m},
\end{align}
where in going from the first to second line we have expanded $xP_k(x)$ again in terms of the monic polynomials. We therefore have that $\braket{xP_n,P_k} = 0$ for $n>k+1$ such that the right hand side of Eq.\ (\ref{Eq:ProofMonicExpansion}) reduces to three terms,
\begin{equation}\label{Eq:PrimalRecRelation}
    xP_n(x) = d_{n+1}P_{n+1}(x) + d_nP_n(x) + d_{n-1}P_{n-1}(x).
\end{equation}
The leading coefficient of $P_n(x)$ is unity by definition such that the same will be the case for $xP_n(x)$, which in turn implies $d_{n+1} = 1$. As for $d_{n-1}$ we can write
\begin{align}
    d_{n-1}\zeta_{n-1} & = \braket{xP_n,P_{n-1}} \\
    & = \braket{P_n,xP_{n-1}} \\
    & = \braket{P_n,P_n + d_{n-1}P_{n-1} + d_{n-2}P_{n-2}} \\
    & = \braket{P_n,P_n},
\end{align}
where we have applied Eq.\ (\ref{Eq:PrimalRecRelation}) when going from the second to third line. Putting all of this together and rearranging the terms of Eq.\ (\ref{Eq:PrimalRecRelation}) we can write
\begin{equation}
    P_{n+1}(x) = \left(x - \frac{\braket{xP_n,P_n}}{\zeta_n}\right)P_n(x) - \frac{\zeta_n}{\zeta_{n-1}}P_{n-1}(x),
\end{equation}
which is the three-term recurrence relation given in Eq.\ (\ref{Eq:ThreeTermRecurrence}).

\section{Transforming the integral representation of the ensemble to a lattice representation}
The Hamiltonian of the entire ensemble continuum can be represented in terms of an integral over all elements of the ensemble as
\begin{equation}\label{Eq:BasicIntegral}
    \hat{H}_{\text{Ens}} = \int\text{d}\lambda \hat{H}_\lambda,
\end{equation}
where $\lambda$ is a label that spans the elements. Rewriting Eq.\ (\ref{Eq:BasicIntegral}) in terms of the $\{\ket{n, \lambda}\}_{n=0}^N$ basis yields
\begin{equation}\label{Eq:IntegralTerms}
    \begin{aligned}
        \hat{H}_{\text{Ens}} = & \sum_{n,m=0}^N\braket{n\vert\hat{H}_0\vert m}\int\text{d}\lambda\ket{n, \lambda}\bra{m, \lambda} + \\
        &\sum_{n,m=0}^N\int\text{d}\lambda f_{n,m}(\lambda)\ket{n, \lambda}\bra{m, \lambda}.
    \end{aligned}
\end{equation}

In order to develop the integral terms in Eq.\ (\ref{Eq:IntegralTerms}) we make an unitary change of basis using a sequence of polynomials $\{p_k(\lambda)\}_{n=0}^\infty$ that are mutually orthonormal with respect to a given probability distribution $p$,
\begin{equation}
    \ket{n, \lambda} = \sum_{k=0}^\infty\sqrt{p(\lambda)}p_k(\lambda)\ket{n, k},
\end{equation}
whose inverse is
\begin{equation}
    \ket{n, k} = \int\text{d}\lambda\sqrt{p(\lambda)}p_k(\lambda)\ket{n, \lambda}.
\end{equation}
We can see that the change of basis is indeed unitary by considering
\begin{align}
    &\braket{n, k \vert n, k^\prime} \nonumber \\
    &= \iint\text{d}\lambda\text{d}\lambda^\prime \sqrt{p(\lambda)p(\lambda^\prime)}p_k(\lambda)p_{k^\prime}(\lambda^\prime)\braket{n, \lambda \vert n, \lambda^\prime} \\
    & = \iint\text{d}\lambda\text{d}\lambda^\prime \sqrt{p(\lambda)p(\lambda^\prime)}p_k(\lambda)p_{k^\prime}(\lambda^\prime)\delta(\lambda - \lambda^\prime) \\
    & = \int\text{d}\lambda p(\lambda)p_k(\lambda)p_{k^\prime}(\lambda) \\
    & = \delta_{k, k^\prime}.
\end{align}

For terms of the form appearing in the first line of Eq.\ (\ref{Eq:IntegralTerms}) we will have
\begin{align}
    &\int\text{d}\lambda\ket{n, \lambda}\bra{m, \lambda} \nonumber \\
    &= \sum_{k,k^\prime=0}^\infty\int\text{d}\lambda p(\lambda)p_k(\lambda)p_{k^\prime}(\lambda)\ket{n, k}\bra{m, k^\prime}  \\
    & = \sum_{k,k^\prime=0}^\infty\braket{p_k,p_{k^\prime}}\ket{n, k}\bra{m, k^\prime} \\
    & = \sum_{k = 0}^\infty\ket{n, k}\bra{m, k}, 
\end{align}
where going from the first to second line we have noted that the integral is the inner product that we had in Eq.\ (\ref{Eq:OrthonormalityCondition}) and then from going from the second to last line we applied the orthogonality of the polynomials. For terms of the form appearing in the second line of Eq.\ (\ref{Eq:IntegralTerms}) we will have
\begin{align}
    & \int\text{d}\lambda f_{n,m}(\lambda)\ket{n, \lambda}\bra{m, \lambda}\nonumber \\
    &= \sum_{k,k^\prime=0}^\infty\int\text{d}\lambda f_{n,m}(\lambda)p(\lambda)p_k(\lambda)p_{k^\prime}(\lambda)\ket{n, k}\bra{m, k^\prime} \\
    & = \sum_{k,k^\prime=0}^\infty f_{n,m}^{(k, k^\prime)}\ket{n, k}\bra{m, k^\prime}
\end{align}
where going from the first to second line we have made the definition $f_{n,m}^{(k, k^\prime)} = \braket{f_{n,m}p_k,p_{k^\prime}}$. We can now rewrite Eq. (\ref{Eq:IntegralTerms}) as a semi-infinite lattice model of the form
\begin{equation}
    \hat{H}_{\text{Ens}} = \sum_{k,k^\prime = 0}^\infty\sum_{n,m=0}^N\left(\braket{n\vert\hat{H}_0\vert m}\delta_{k,k^\prime} + f_{n,m}^{(k,k^\prime)}\right)\ket{n, k}\bra{m, k^\prime},
\end{equation}
where the $\lambda$-independent component of the ensemble has been projected onto each node of the lattice, whilst the $\lambda$-dependent component is responsible for the $f_{n,m}^{(k,k^\prime)}$ lattice coupling terms. 

\subsection{Linear disorder}
At this point it is worth noting that we have transformed the integral representation of $\hat{H}_{\text{Ens}}$ into a lattice model with respect to an \textit{arbitrary} probability distribution with well defined moments, and that the utility of the lattice mapping will depend on whether or not our choice of $p$ will result in a simple form for the $f_{n,m}^{(k,k^\prime)}$ coupling constants. We now demonstrate how such a choice is simple to determine for most models of disorder that are considered in the literature by exploiting the recurrence relation that orthogonal polynomials satisfy. 

Let us assume that $f$ is a linear function of disorder such that $f(\lambda) = w\lambda$, where $w$ has units of energy and can be considered to characterize the strength of the disorder contribution. We can then compute the coupling coefficients of the lattice expansion using the three-term recurrence relation,
\begin{align}
    f^{(k,k^\prime)} & = w\braket{\lambda p_k,p_{k^\prime}} \\
    & = \frac{w}{\sqrt{\zeta_k\zeta_{k^\prime}}}\braket{\lambda P_k,P_{k^\prime}} \\
    & = w\left(\sqrt{\beta_{k+1}}\delta_{k+1,k^\prime} + \alpha_k\delta_{k,k^\prime} + \sqrt{\beta_{k}}\delta_{k-1,k^\prime}\right).
\end{align}
We can see that the resulting interaction terms in the lattice will be of the nearest-neighbour form, and the strength of this coupling is determined by the square root of the recurrence coefficients, $\sqrt{\beta_n}$. We also have terms corresponding to energy shifts that are given by $\alpha_n$. Finally note that $\alpha_n = \alpha$ for all $n$ if the probability distribution is even with respect to a point $\alpha$, i.e. $p(\alpha-\lambda) = p(\alpha+\lambda)$. This can be seen by recalling their definition given in Eq.\ (\ref{Eq:alpha_n_def}) and noting that the resulting polynomials will have an even and odd parity with respect to the point $\alpha$. 

\subsection{Example: Two-level system ensemble}
Let us consider an ensemble of non-interacting qubits that have their excited state energies to be randomly sampled from a probability distribution $p(\lambda)$, such that the dynamics of each realization is determined by a Hamiltonian of the form
\begin{equation}
    \hat{H}_\lambda = E_0\ket{0, \lambda}\bra{0, \lambda} + (E_1 + w\lambda)\ket{1, \lambda}\bra{1, \lambda},
\end{equation}
where $E_0$ and $E_1$ are the central energies of the ground and excited state, respectively, $\lambda$ is drawn from a random distribution $p(\lambda)$, and $w$ characterizes the width of the static disorder. The Hamiltonian of the ensemble is then mapped onto a nearest-neighbour coupled chain form,
\begin{equation}
    \begin{aligned}
        \hat{H}_{\text{Ens}} &= \sum_{k=0}^\infty\left(E_0\ket{0, k}\bra{0, k} + (E_1 + w\alpha_k)\ket{1, k}\bra{1, k}\right) \\
        &+ w\sum_{k=0}^\infty\sqrt{\beta_{k+1}}\ket{1, k}\bra{1, k+1} + \text{h.c.}
    \end{aligned}
\end{equation}
where $\alpha_k$ and $\beta_k$ are the recurrence coefficients stemming from performing the lattice expansion with respect to the distribution of the disorder, $p(\lambda)$. We take each realization to have an initial state of the form $\ket{\psi_0}_\lambda = a\ket{0, \lambda} + b\ket{1, \lambda}$, with $\vert a\vert^2 + \vert b\vert^2 =1$, which will then yield an initial state of the ensemble in terms of the lattice basis that is of the form $\ket{\Psi_0} = a\ket{0, 0} + b\ket{1, 0}$. We can now perform a partial trace over all nodes of the lattice in order to compute the ensemble averaged dynamics,
\begin{align}
    \Bar{\rho}(t) &= \sum_{k=0}^\infty\braket{k\vert \rho(t)\vert k} \\
    &= \sum_{n,m = 0, 1} \left[\sum_{k=0}^\infty\braket{n, k\vert \rho(t)\vert m, k}\right] \ket{n}\bra{m},
\end{align}
where $\rho(t) = e^{-i\hat{H}_\text{Ens}t}\rho_0e^{i\hat{H}_\text{Ens}t}$ and $\rho_0 = \vert a\vert^2\ket{0, 0}\bra{0, 0} + \vert b\vert^2\ket{1, 0}\bra{1, 0} + ab^*\ket{0, 0}\bra{1, 0} + a^*b\ket{1, 0}\bra{0, 0}$. 

We define functions $c_{n,m}(t) = \sum_{k=0}^\infty\braket{n, k\vert \rho(t)\vert m, k}$ for the sake of brevity and compute the averaged dynamics for this example. We can immediately determine $c_{0,0}(t) = \vert a\vert^2$ by noting that $\ket{0, 0}$ is a stationary state of $\hat{H}_\text{Ens}$. We can reduce $c_{0,1}(t)$ to a single term,
\begin{align}
    c_{0,1}(t) & = ab^*\sum_{k=0}^\infty \braket{0, k\vert e^{-i\hat{H}_\text{Ens}t}\vert 0, 0}\braket{1, 0\vert e^{i\hat{H}_\text{Ens}t}\vert 1, k} \\
    & = ab^* e^{-iE_0t}\braket{1, 0\vert e^{i\hat{H}_\text{Ens}t}\vert 1, 0},
\end{align}
and then note that $\braket{1, 0\vert e^{i\hat{H}_\text{Ens}t}\vert 1, 0}$ is the \textit{characteristic function} of the random variable $\Lambda = w\lambda + E_1$ \cite{Lukacs1972Apr}. 
We denote the characteristic function of a random variable $X$ as $\varphi_X$ such that we may now write
\begin{equation}
    c_{0,1}(t) = ab^* e^{-i(E_0-E_1)t}\varphi_{w\lambda}(t),
\end{equation}
where we have made use of the fact that $\varphi_{\Lambda}(t) = e^{iE_1t}\varphi_{w\lambda}(t)$. As $c_{1,0}(t) = c_{0,1}^*(t)$, we now only need to determine $c_{1,1}(t)$, which will reduce as
\begin{equation}\label{Eq:c_11_equation}
    c_{1,1}(t) = \vert b\vert^2 \sum_{k=0}^\infty\left\vert\braket{1, k\vert e^{-i\hat{H}_\text{Ens}t}\vert 1, 0}\right\vert^2.
\end{equation}
We might already expect $c_{1,1}(t)$ to be constant for all times as we know that populations of the $\ket{0, \lambda}$, $\ket{1, \lambda}$ states are conserved quantities. It is however not so obvious that this should be the case when considering Eq.\ (\ref{Eq:c_11_equation}). 

Notice that we can write the $\ket{1, k}$ elements in terms of $\ket{1, 0}$ in a compact form if we allow for the corresponding orthonormal polynomials to take operator arguments, i.e. we have
\begin{equation}
    \ket{1, k} = p_k(\hat{H}_{\text{Ens}})\ket{1, 0}.
\end{equation}
We then use this fact in conjunction to $\hat{H}_{\text{Ens}}e^{-i\hat{H}_\text{Ens}t} = i\partial_t e^{-i\hat{H}_\text{Ens}t}$ in order to write
\begin{align}
    \braket{1, k\vert e^{-i\hat{H}_\text{Ens}t}\vert 1, 0} & = \braket{1, 0\vert p_k(\hat{H}_{\text{Ens}})e^{-i\hat{H}_\text{Ens}t}\vert 1, 0} \\
    & = p_k(i\partial_t)\braket{1, 0\vert e^{-i\hat{H}_\text{Ens}t}\vert 1, 0} \\
    & = p_k(i\partial_t)\varphi_{\Lambda}^*(t),
\end{align}
and in general we have
\begin{equation}
    \braket{1, k\vert e^{i\hat{H}_\text{Ens}t}\vert 1, k^\prime} = p_k(-i\partial_t)p_{k^\prime}(-i\partial_t)\varphi_{\Lambda}(t).
\end{equation}
We can now rewrite Eq.\ (\ref{Eq:c_11_equation}) in terms of the characteristic function as
\begin{equation}
    c_{1,1}(t) = \vert b\vert^2 \sum_{k=0}^\infty\left\vert p_k(i\partial_t)\varphi_{\Lambda}^*(t)\right\vert^2.
\end{equation}
In this form we can now readily consider $i\partial_t c_{1,1}(t)$ and apply the three-term recurrence relation for $P_k$ in order to see that $c_{1,1}(t)$ is indeed a constant. We have
\begin{widetext}
    \begin{align}
        i\partial_t c_{1,1}(t) & = \vert b\vert^2 \sum_{k=0}^\infty\frac{1}{\zeta_k}\left(\left[i\partial_tP_k(i\partial_t)\varphi_{\Lambda}^*(t)\right]P_k(-i\partial_t)\varphi_{\Lambda}(t) - P_k(i\partial_t)\varphi_{\Lambda}^*(t)\left[-i\partial_tP_k(-i\partial_t)\varphi_{\Lambda}(t)\right]\right) \\
        & = \vert b\vert^2 \sum_{k=0}^\infty\frac{1}{\zeta_k}\left(\left[P_{k+1}(i\partial_t) + \alpha_kP_k(i\partial_t) + \beta_kP_{k-1}(i\partial_t)\right]\varphi_{\Lambda}^*(t)P_k(-i\partial_t)\varphi_{\Lambda}(t) \right.\nonumber\\
        & \left.\hspace{1.9cm}- P_k(i\partial_t)\varphi_{\Lambda}^*(t)\left[P_{k+1}(-i\partial_t) + \alpha_kP_k(-i\partial_t) + \beta_kP_{k-1}(-i\partial_t)\right]\varphi_{\Lambda}(t)\right) \\
        & = 2\vert b\vert^2\sum_{k=0}^\infty\text{Im}\left( \frac{1}{\zeta_k}P_{k+1}(i\partial_t)\varphi_{\Lambda}^*(t)P_k(-i\partial_t)\varphi_{\Lambda}(t) - \frac{1}{\zeta_{k-1}}P_k(i\partial_t)\varphi_{\Lambda}^*(t)P_{k-1}(-i\partial_t)\varphi_{\Lambda}(t)\right) \\
        & = 2\vert b\vert^2\lim_{k\to\infty}\frac{1}{\zeta_k}\text{Im}\left(P_{k+1}(i\partial_t)\varphi_{\Lambda}^*(t)P_k(-i\partial_t)\varphi_{\Lambda}(t)\right),
    \end{align}
\end{widetext}
whereby having collected like-terms we have noted that we have an infinite series of the form $\sum_{k=0}^\infty a_k - a_{k-1} = \lim_{k\to\infty}a_k$ (with $a_{-1} = 0$), so we are left with a single term of the form $\frac{1}{\zeta_k}P_{k+1}(i\partial_t)\varphi_{\Lambda}^*(t)P_k(-i\partial_t)\varphi_{\Lambda}(t) = \sqrt{\beta_{k+1}}\braket{1, k+1\vert e^{-i\hat{H}_\text{Ens}t}\vert 1, 0}\braket{1, 0\vert e^{i\hat{H}_\text{Ens}t}\vert 1, k}$ which goes to zero as $k\to\infty$ for a local Hamiltonian at finite times.

Combining the fact that $\partial_t c_{1,1}(t) = 0$ with the initial condition $c_{1,1}(0) = \vert b\vert^2$ we have that $c_{1,1}(t) = \vert b\vert^2$. We therefore have the ensemble averaged dynamics to be
\begin{equation}
    \begin{aligned}
        \Bar{\rho}(t) = & \vert a\vert^2\ket{0}\bra{0} + \vert b\vert^2\ket{1}\bra{1} \\
        & + ab^*e^{-i(E_0 - E_1)t}\varphi_{w\lambda}(t)\ket{0}\bra{1} + \text{h.c.}
    \end{aligned}
\end{equation}
which is in agreement with what has been derived in \cite{Kropf2016Aug}.

\subsection{Example: Disorder dependent initial state}
The pure state formalism is able to treat physical situations where the initial state of each realization is correlated to the disorder. We illustrate this point here with an example of an ensemble of systems, $\hat{H}_\lambda$, where $\lambda$ follows a Wigner semicircle distribution $p(\lambda) = \frac{2}{\pi}\sqrt{1-\lambda^2}$ such that $\lambda$ can take values on the range $[-1, 1]$. We shall consider the initial state of each realization to be of the form
\begin{equation}
    \ket{\psi_0, \lambda} = \lambda\ket{0,\lambda} + \sqrt{1-\lambda^2}\ket{1,\lambda}
\end{equation}
The initial state of the ensemble is now written as
\begin{align}
    \ket{\Psi_0} &= \int_{-1}^{1}\text{d}\lambda \sqrt{p(\lambda)}\left(\lambda\ket{0,\lambda} + \sqrt{1-\lambda^2}\ket{1,\lambda}\right) \\
    &= \sum_{k=0}^\infty d_{0,k}\ket{0,k} + \sum_{k=0}^\infty d_{1,k}\ket{1,k},
\end{align}
with coefficients
\begin{align}
    d_{0,k} &= \int_{-1}^{1}\text{d}\lambda p(\lambda)U_k(\lambda)\lambda, \\
    d_{1,k} &= \int_{-1}^{1}\text{d}\lambda p(\lambda)U_k(\lambda)\sqrt{1-\lambda^2},
\end{align}
where $U_k(\lambda)$ are the Chebyshev polynomials of the second kind \cite{Boyd, Abramowitz1998Jun}. Computing these integrals yields
\begin{align}
    d_{0,k} &= \frac{1}{2}\delta_{1,k}, \\
    d_{1,k} &= -\frac{4}{\pi}\frac{1+\cos(\pi k)}{(k+3)(k^2-1)},
\end{align}
such that our initial ensemble state is discretized as
\begin{equation}
    \ket{\Psi_0} = \frac{1}{2}\ket{0,1} - \frac{8}{\pi}\sum_{k=0}^\infty \frac{1}{(2k+3)(4k^2-1)}\ket{1,2k}.
\end{equation}

\begin{figure}
    \centering
    \includegraphics[width=\linewidth]{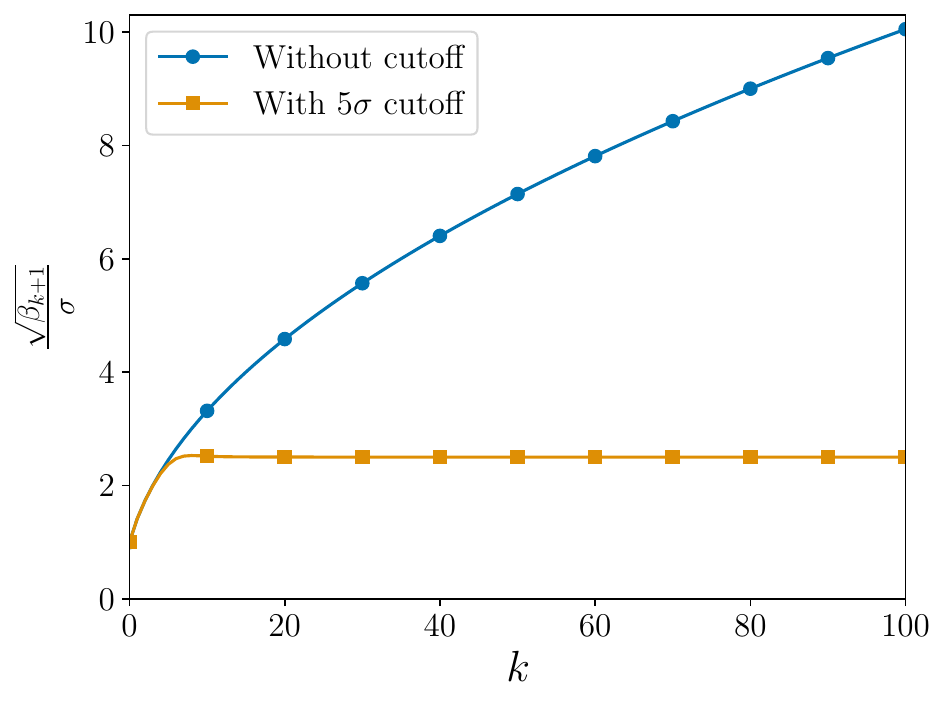}
    \caption{Lattice coupling strength behaviour for an ensemble with Gaussian disorder, with and without an energy cutoff at $5\sigma$. We are able to simulate dynamics for longer times by introducing an energy cutoff.}
    \label{Fig:couplings_w_wo_cutoff}
\end{figure}

\section{Introducing an energy cutoff}
The mapping breaks down for cases of disorder which follow a distribution whose moments are undefined, e.g. Cauchy and Lévy distributions, which then in turn result in undefined recurrence coefficients. The divergence implicit in the integral form of $\hat{H}_{\text{Ens}}$ is made explicit in its lattice form. We can remedy the situation by introducing an energy cutoff which then leads to well defined recurrence coefficients and recover approximate averaged dynamics whose accuracy depends on the choice of cutoff.

It is also useful to apply an energy cutoff even in cases where all moments of the distribution are well defined, as doing so will ensure that the recurrence coefficients will reach asymptotic values. Take for example a Gaussian distribution with mean zero and standard deviation $\sigma$ such that the recurrence coefficients will be $\alpha_k = 0$ for all $k$ and $\beta_{k+1} = \sigma^2(k+1)$. Due to the strictly increasing value of the lattice couplings we find that there are diminishing returns in increasing the lattice truncation, making it more and more computationally intensive to simulate dynamics to longer times. A $5\sigma$ cutoff will lead to the couplings having an asymptotic value of $\lim_{k\to\infty}\sqrt{\beta_{k+1}} = \frac{5}{2}\sigma$, thus curbing the monotonic increase of the lattice couplings whilst simultaneously preserving the fidelity of the distribution up to $\pm 5\sigma$. Figure \ref{Fig:couplings_w_wo_cutoff} demonstrates this behaviour of the lattice couplings corresponding to a Gaussian disorder distribution with and without an energy cutoff.

\begin{figure}
    \centering
    \includegraphics[width=\linewidth]{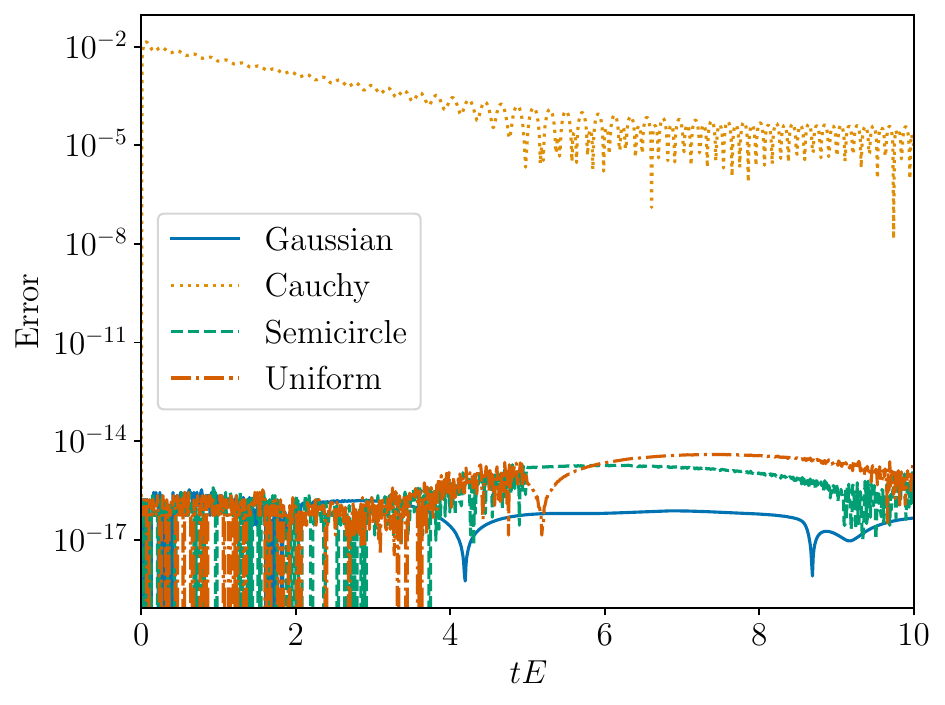}
    \caption{Disordered qubit ensemble dephasing dynamics error.}
    \label{Fig:QubitError}
\end{figure}

\section{Table of quantities}
\begin{center}
    \def\arraystretch{1.5}
    \begin{tabular}{ |c|c|c|c|c|c| } 
         \hline
         Name & $p(\lambda)$ & $\mathcal{D}$ & $\alpha_k$ & $\beta_{k+1}$ & $\varphi_{\Lambda}(t)$ \\ 
        \hline
        \hline
         Gaussian & $\frac{1}{\sqrt{2\pi}\sigma}e^{-\lambda^2 / 2\sigma^2}$ & $\mathbb{R}$ & 0 & $\sigma^2(k+1)$ & $e^{-\sigma^2t^2/2}$ \\ 
        \hline
         Cauchy & $\frac{1}{\pi\theta}\frac{\theta^2}{\lambda^2 + \theta^2}$ & $\mathbb{R}$ & N/A & N/A & $e^{-\theta |t|}$ \\ 
        \hline
         Semicircle & $\frac{2}{\pi w^2}\sqrt{w^2 - \lambda^2}$ & $[-w,w]$ & 0 & $\frac{w}{2}$ & $2\frac{J_1(wt)}{wt}$ \\
        \hline
         Uniform & $\frac{1}{2v}$ & $[-v,v]$ & 0 & $\frac{v(k+1)}{\sqrt{4(k+1)^2 - 1}}$ & $\frac{\sin(vt)}{vt}$ \\
         \hline
    \end{tabular}
\end{center}

For the qubit dephasing considered in the main text we have set $\sigma = \theta = w = v = E_1 - E_0$. An energy cutoff of $E_\pm = \pm30\theta$ is introduced for the Cauchy distribution in order to achieve well defined recurrence coefficients. 


\bibliographystyle{apsrev4-1}
\bibliography{mybib}

\end{document}